\documentclass[a4paper,fleqn]{cas-dc}
\usepackage[utf8]{inputenc}
\usepackage{newunicodechar}
\usepackage{graphicx}
\usepackage{subcaption}  
\newunicodechar{♫}{\textmusicalnote}
\usepackage[numbers]{natbib}

\def\tsc#1{\csdef{#1}{\textsc{\lowercase{#1}}\xspace}}
\tsc{WGM}
\tsc{QE}
\tsc{EP}
\tsc{PMS}
\tsc{BEC}
\tsc{DE}

\begin{document}
\let\WriteBookmarks\relax
\def\floatpagepagefraction{1}
\def\textpagefraction{.001}
\shorttitle{Leveraging social media news}
\shortauthors{Q. Wei et~al.}

\title [mode = title]{A Query-Aware Multi-Path Knowledge Graph Fusion Approach for Enhancing Retrieval-Augmented Generation in Large Language Models}                   
\author[1]{Qikai Wei}[orcid=0000-0002-1048-3814]
\ead{weiqikai@xs.ustb.edu.cn}

\author[1]{Huansheng Ning}[orcid=0000-0001-6413-193X]
\ead{ninghuansheng@ustb.edu.cn}
\cormark[1]

\author[1]{Chunlong Han}[orcid=0000-0003-1602-1597]
\ead{hanchunlong@xs.ustb.edu.cn}

\author[2]{Jianguo Ding}[orcid=0000-0002-8927-0968] 
\ead{jianguo.ding@bth.se}

\affiliation[1]{
  organization={School of Computer and Communication Engineering, University of Science and Technology Beijing},
  city={Beijing},
  country={China}
}

\affiliation[2]{
  organization={Blekinge Institute of Technology},
  city={Karlskrona},
  country={Sweden}
}

\cortext[cor1]{Corresponding author}

\begin{abstract}
Retrieval Augmented Generation (RAG) has gradually emerged as a promising paradigm for enhancing the accuracy and factual consistency of content generated by large language models (LLMs). However, existing RAG studies primarily focus on retrieving isolated segments using similarity-based matching methods, while overlooking the intrinsic connections between them. This limitation hampers performance in RAG tasks. To address this, we propose QMKGF, a Query-Aware Multi-Path Knowledge Graph Fusion Approach for Enhancing Retrieval Augmented Generation. First, we design prompt templates and employ general-purpose LLMs to extract entities and relations, thereby generating a knowledge graph (KG) efficiently. Based on the constructed KG, we introduce a multi-path subgraph construction strategy that incorporates one-hop relations, multi-hop relations, and importance-based relations, aiming to improve the semantic relevance between the retrieved documents and the user query. Subsequently, we designed a query-aware attention reward model that scores subgraph triples based on their semantic relevance to the query. Then, we select the highest score subgraph and enrich subgraph with additional triples from other subgraphs that are highly semantically relevant to the query. Finally, the entities, relations, and triples within the updated subgraph are utilised to expand the original query, thereby enhancing its semantic representation and improving the quality of LLMs' generation. We evaluate QMKGF on the SQuAD, IIRC, Culture, HotpotQA, and MuSiQue datasets. On the HotpotQA dataset, our method achieves a ROUGE-1 score of 64.98\%, surpassing the BGE-Rerank approach by 9.72 percentage points (from 55.26\% to 64.98\%). Experimental results demonstrate the effectiveness and superiority of the QMKGF approach.

\end{abstract}



\maketitle
\section{Introduction}
In recent years, Large Language Models (LLMs) have demonstrated remarkable performance in the field of Natural Language Processing (NLP), finding widespread application across various artificial intelligence tasks \cite{gpt, wu2024effective, zhu2025legn, XU2025127582}. Despite their increasingly powerful generative capabilities, LLMs often produce outputs that appear plausible but are factually incorrect—a phenomenon commonly referred to as hallucination \cite{fu2025cue, hallucinations}.

To address this issue, Retrieval-Augmented Generation (RAG) has emerged as an effective solution \cite{gao2023retrieval, zhao2024retrieval}. By incorporating external knowledge sources to provide contextual support for generation, RAG significantly mitigates the hallucination phenomenon commonly observed in LLMs and substantially enhances the factual accuracy and reliability of the generated content \cite{yu2024evaluation}. Owing to its ability to improve factual consistency, RAG has been widely applied in various domains such as law \cite{Chatlaw}, medical \cite{medical_graph_rag}, and tourism \cite{cultour}. Fig. 1 illustrates a comparison between RAG-based approaches and the standard LLM-only paradigm. The green line indicates the LLM-only setting, where the model generates answers directly from the query without any external retrieval. The red line represents the RAG-based approach, where a user query is used to retrieve relevant chunks from a document corpus, guiding the LLM to generate more informed responses. The blue line denotes the final response delivered to the user by the LLM.

\begin{figure}[t]
	\begin{center}
		\includegraphics[scale=0.8]{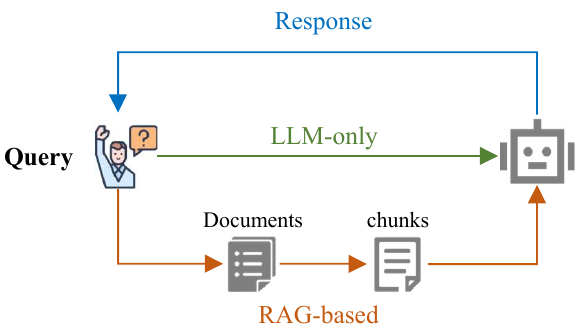}
		\caption{Comparison between LLM-only and RAG-based paradigm.}
		\label{fig:1}       
	\end{center}
\end{figure}

Existing RAG methods can be divided into document retrieval-based and graph structure-based methods according to the organization of knowledge structure. Document retrieval-based RAG commonly adopt query expansion \cite{Query2doc} or re-ranking mechanisms to improve the precision and recall of relevant content. For example, NetEase Youdao \cite{BCEmbedding} proposed the BCEmbedding method, which employs a two-stage approach: the first stage performs efficient initial retrieval using embeddings, while the second stage applies a RM to conduct fine-grained semantic re-ranking of the retrieved results, thereby enhancing the overall retrieval quality. Xiao et al. \cite{C-pack} introduced the BGE-Rerank model to re-rank the top-k documents retrieved by the embedding model, aiming to enhance the relevance between the retrieved documents and the query. This refinement ultimately boosts the performance of RAG tasks. However, these methods heavily rely on similarity-based query-to-chunk matching, which may introduce irrelevant content and compromise the accuracy of the response. While such approaches are effective for simple tasks, they often fall short when dealing with complex queries \cite{jiang2023active}.

In contrast, graph graph structure-based retrieval methods for RAG leverage graph augmentation or path-guided mechanisms to enhance the semantic relevance of retrieval results. Representative works include GraphRAG \cite{graphrag}, QCG-rerank \cite{QCG-Rerank}, and KG2RAG \cite{kg2rag}, which improve content retrieval quality by integrating the semantic relationships between entities. However, existing graph-based RAG approaches typically rely on a single path, which limits the scope of retrieved content and affects the diversity of the final generated output. On the one hand, current methods often construct subgraphs based on fixed rules or structural distances, neglecting the variation in information-bearing capacity across different semantic paths. This makes it difficult to balance semantic diversity and path significance, resulting in the introduction of fragmented and structurally homogeneous information during retrieval. On the other hand, incorporating multiple paths directly may lead to semantic redundancy and the accumulation of noise, which weakens the model's ability to identify critical information. Therefore, if a mechanism can be developed that not only extracts high-quality information from multiple paths but also enables semantic aggregation and filtering, it holds the potential to improve generation quality while mitigating redundancy and noise in the process of graph information incorporation.

To address the above challenges, we propose QMKGF, a Query-aware Multi-path Knowledge Graph Fusion approach. This method enhances the relevance of retrieved content and the quality of generated responses in the RAG framework by jointly leveraging the structure of knowledge graphs and the semantics of the query. Specifically, entities in a knowledge graph often have multiple adjacent paths, each representing different potential semantic associations. We begin by extracting key entities from the input query and use them as anchors to perform multi-path semantic expansion within the knowledge graph, thereby retrieving information that is more helpful for answering the original query. Considering the heterogeneity of semantic relations in different paths—such as one-hop, multi-hop, or importance-based paths—we design a multi-path subgraph construction mechanism to comprehensively capture relevant semantic evidence. However, many paths or triples may lack precise semantic alignment with the query, potentially introducing irrelevant or noisy information into the retrieval process. To address this, we introduce a query-aware attention reward model, which performs fine-grained scoring of triples across different paths based on their semantic relevance to the query. Compared to traditional vector similarity-based methods, our attention model exhibits stronger query-awareness, thereby reducing noise introduced by semantic mismatches. We then select the most query-relevant subgraph as the backbone and fuse high-relevance triples from other paths to form a compact, semantically aligned fused subgraph. This fused subgraph is subsequently used to expand the query, further improving the relevance of document retrieval and ultimately enhancing the response quality of LLMs. We evaluate QMKGF on five benchmark datasets: SQuAD, IIRC, Culture, HotpotQA, and Musique. On the HotpotQA dataset, our method achieves a ROUGE-1 score of 64.98\%, surpassing the BGE-Rerank approach by 9.72 percentage points (from 55.26\% to 64.98\%). Experimental results demonstrate that QMKGF significantly outperforms existing methods and shows clear advantages in improving answer accuracy.

In summary, the contributions of our paper are as follows:

1. A multi-path KG subgraph construction method is developed, incorporating one-hop relations, multi-hop relations, and importance-based relations to capture diverse and salient knowledge paths.

2. A query-aware attention reward model is introduced, enabling fine-grained scoring of subgraph triples based on their semantic relevance to the input query.

3. A subgraph fusion strategy is proposed, which selects the highest-scoring subgraph and integrates semantically relevant triples from other subgraphs, resulting in a more informative and query-aligned KG subgraph.

4. A KG-based query expansion approach is presented, enriching the original query with entities, relations, and triples from the final subgraph to improve the quality and factual consistency of RAG outputs.

\section{Related Work}
In recent years, Large Language Models (LLMs) have shown impressive performance in both comprehension and text generation, especially within the context of open-domain QA tasks\cite{gpt, llama}. They have been widely applied in areas such as QA systems \cite{cultour, li2025two, xiao2024tpke}, code generation \cite{PATSAKIS2024124912,nie2021coregen,BAI2025126357}. However, LLMs' heavy reliance on static knowledge embedded in their parameters. When applied to scenarios characterized by rapidly evolving information and high factual requirements, LLMs are prone to producing outdated or factually inaccurate responses. This limitation frequently leads to hallucinations, thereby undermining the overall quality and trustworthiness of the generated outputs \cite{hallucinations,zhang2025lgkgr}. To mitigate the aforementioned limitations, Retrieval Augmented Generation (RAG) is a technique that combines information retrieval with text generation, aiming to improve the accuracy of generated content and reduce hallucinations by incorporating support from external documents \cite{guu2020retrieval}. The typical workflow involves two stages: first, a retrieval model is used to fetch relevant information from external knowledge sources based on the user's query; then, this retrieved content is input into LLMs to integrate the information and generate a response.

\textbf{RAG for LLMs}

RAG has emerged as a mainstream approach for mitigating hallucination issues in LLMs due to its ability to retrieve relevant factual content. By introducing external factual knowledge during the inference process, RAG provides contextual support for LLMs, thereby enhancing the factual consistency and reliability of the generated content \cite{Chatlaw, peng2023check, xu2023search}. Existing LLM-based RAG methods can be broadly categorized into two paradigms: document retrieval-based RAG and graph retrieval-based RAG.

For document retrieval–based RAG, embedding serves as a critical component. It determines the ability to vectorize text, bringing semantically similar content closer in the vector space. The common method is to divide long text into several chunks, which weakens the semantic integrity and context coherence. To address this, Luo et al. \cite{luo2024bge} proposed Landmark Embedding, which employs a sliding-window embedding training method that preserves contextual consistency and enhances the model’s ability to represent extended text. Additionally, to enrich the semantic content of user queries, Wang et al. \cite{Query2doc} introduced a method that uses a few-shot prompting strategy to guide LLMs in generating pseudo-documents. These are concatenated with the original queries to form augmented queries, allowing the retrieval process to incorporate both the foundational knowledge of the LLMs and the content of external documents. Shi et al. \cite{shi2024generate} proposed Genground, a model that first uses the query as input to an LLM to generate a base answer, and then refines potential errors in this answer using retrieved content.

\begin{figure*}[ht]
	\begin{center}
		\includegraphics[scale=0.6]{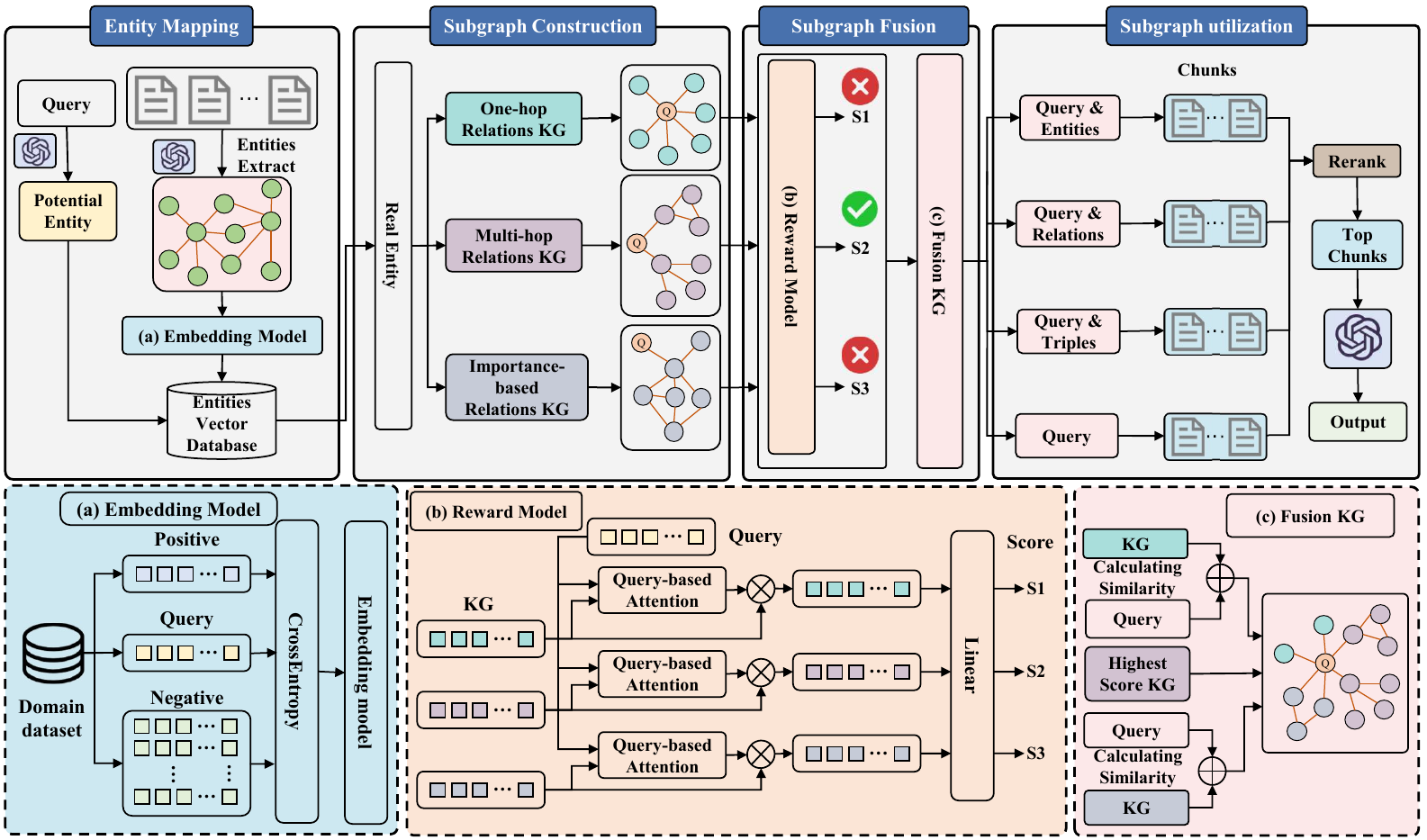}
		\caption{Framework of the proposed QMKGF.}
		\label{fig:2}       
	\end{center}
\end{figure*}

For graph retrieval-based RAG methods, graphs are typically integrated with the RAG framework to enhance the model's generalization capability \cite{li2024graph}. Li et al. \cite{graphrag} proposed GraphRAG, which constructs semantic graphs to enable query-focused summarization, thereby improving the quality of responses to user queries. Zhu et al. \cite{kg2rag} proposed KG2RAG, a framework that integrates structured knowledge from knowledge graphs to reinforce the semantic connections between document chunks, thereby improving both retrieval accuracy and generation quality. In the cultural tourism domain, Wei et al. \cite{QCG-Rerank} proposed QCG-rerank, which constructs a query-document graph structure and integrates query expansion with reranking mechanisms to enhance the relevance and expressiveness of retrieved passages.

Existing RAG methods are constrained by similarity-based matching between queries and chunks, often overlooking entities in the query and their deep semantic associations across chunks. To address this issue, we propose QMKGF, an approach that automatically constructs KGs. For the entities in the query, we first employ a multi-path KG subgraph construction strategy based on one-hop relations, multi-hop relations, and importance-based relations. This approach enriches the relevance of entity information within the query, thereby enhancing the relevance of retrieved content during the recall process and ultimately improving the question-answering capabilities of LLMs.

\section{Methodology}
In this section, we provide a detailed introduction to QMKGF, whose overall architecture is illustrated in Fig. 2. The model consists of four main components: entity mapping, subgraph construction, subgraph fusion, and subgraph utilization. 1) For entity mapping, we first design prompts to leverage LLMs for KG extraction and construct an entity vector database. Then, potential entities are extracted from the query and mapped to real entities in the KG by matching them against the entity vector database (Section 3.1). 2) For subgraph construction, we propose a multi-path subgraph generation strategy that incorporates one-hop relations, multi-hop relations, and importance-based relations (Section 3.2). 3) For subgraph fusion, a query-aware attention reward model is used to score the three generated subgraphs. We then integrate valid triples from the lower-scoring subgraphs into the higher-scoring one to produce the final subgraph (Section 3.3). 4) In the subgraph utilization module, entities, relations, and triples from the final subgraph are concatenated with the original query for retrieval over a vector database. Relevant document chunks are retrieved and re-ranked using a reranking module, after which the top-ranked chunks are fed into the LLMs to generate the final response (Section 3.4).

\subsection{Entity Mapping}
In the context of processing a large collection of unstructured documents, denoted as $D$ = \{$d_1$,$d_2$, ..., $d_n$\}, we first leverage LLMs to extract entities and relations from the unstructured texts and construct a KG accordingly. The prompt design details for this construction are illustrated in Fig. 1. 

Subsequently, we refined the embedding representations to improve their capacity for semantic comprehension and entity correspondence. The training dataset was constructed in the format \{"query": str, "pos": List[str], "neg": List[str]\}, where "pos" refers to the relevant answers, and "neg" includes irrelevant samples randomly drawn from other documents. This corpus was employed to adapt the embedding model to the target domain through fine-tuning. The optimized loss function is as follows:
\begin{equation}
    \mathcal{L}_i = -\log \frac{\exp\left(\text{sim}(q_i, p_i)/m\right)}{\sum_{j} \exp\left(\text{sim}(q_i, d_j)/m\right)}
\end{equation}
where, $q_i$ denotes the embedding vector of the $i$-th query, $p_i$ denotes positive docment of the $i$-th query, $d_j$ denotes all passages, including both positive and negative examples. $m$ denotes the temperature scaling parameter. sim() denotes the similarity calculation function.

Subsequently, we utilize the fine-tuned embedding model to vectorize the entities in the KG, resulting in entities vector database $\mathit{ent}\_{VB}$ to support entity matching in the subsequent retrieval process. Through the above steps, we ultimately obtain a complete KG G=(V,E) and entities vector database $\mathit{ent}\_{VB}$. 

\begin{figure}[t]
	\begin{center}
		\includegraphics[scale=0.7]{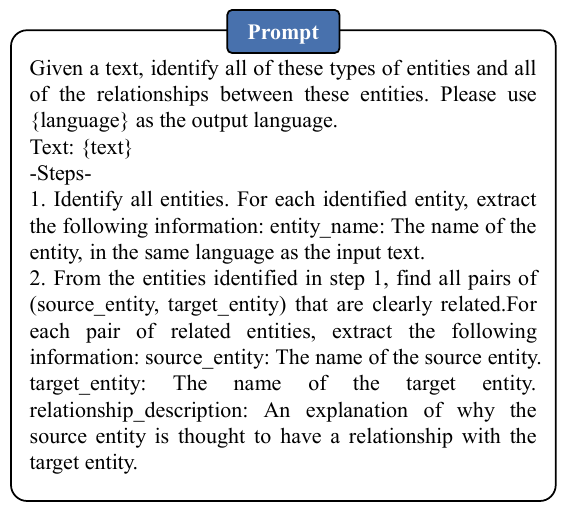}
		\caption{Prompt templates for entities and relations extraction.}
		\label{fig:4}       
	\end{center}
\end{figure}

Given a query $q$, the entity mapping module first employs predefined prompt templates to extract potential entities $e_p$ from the original query. Then, we calculate the similarity between $e_p$ and each entity $e_i \in \mathit{ent}\_{VB}$ to identify the most semantically similar entity. Through this process, the module maps $e_p$ to the most relevant existing entity $e_t \in \mathit{ent}\_{VB}$ within the KG.

\begin{equation}
\mathrm{sim}(e_p, e_i) = \frac{e_p \cdot e_i}{\lVert e_p \rVert \lVert e_i \rVert}, \quad \forall e_i \in \mathit{ent}\_{VB}
\end{equation}

\subsection{Subgraph Construction}
To fully leverage the contextual information associated with each entity, we design a multi-path subgraph construction algorithm. This algorithm integrates one-hop relations, multi-hop relations, and community-level entities based on personalized PageRank. The overall process is illustrated in Fig. 1.

\textbf{One-hop relations}: Given the identified truth entity $e_t$, we first construct its one-hop neighbor set $\mathcal{N}_1(e_t)$, defined as $\left\{ v \in V \mid (e_t, v) \in E \right\}$. Within this set, we compute the semantic similarity between $e_t$ and each neighboring entity $e_i \in \mathcal{N}_1(e_t)$, and select the top-$K$ most relevant nodes. The resulting one-hop subgraph is defined as:
\begin{equation}
\mathrm{Subkg}_{\text{onehop}} = \mathrm{top}_K\left( \mathrm{sim}(e_t, e_i) \right), \quad e_i \in \mathcal{N}_1(e_t)
\end{equation}
where $\mathrm{sim}(\cdot)$ denotes the similarity computation function, and $\mathrm{top}(\cdot)$ refers to selecting the top-$K$ entities from a set. We select $K$ one-hop neighbors of the target entity to enrich its representation by leveraging the contextual information provided by its immediate neighbors.

\textbf{Multi-hop relations}: We select the two most relevant entities $e_o \in N_1(e_t)$ from the one-hop neighbors to perform multi-hop entity expansion. Specifically, we construct a multi-hop entity set $NM$ = \{ $v$ $\in$ $V$ $\mid$ ($e_o$, $v$) $\in$ $E$ \},  and then select the top-$K$ most relevant nodes from this set to obtain the multi-hop subgraph, denoted as $\text{Subkg}_{\text{mulhop}}$.

\textbf{Importance-based relations}: In the importance scoring module, we employ the PageRank algorithm to quantify the relative importance of different components within the community. We first construct a personalized vector $p$ in accordance with Eq. (4).
\begin{equation}
p(v) = 
\begin{cases} 
1, & \text{if } v = e_t \\
0, & \text{otherwise}
\end{cases}
\end{equation}

By iteratively applying Eq. (5), we compute the PageRank score $S(v)$ for each node $v$:
\begin{equation}
S^{(t+1)}(v) = (1 - d) \cdot p(v) + d \sum_{u \in \text{In}(v)} \frac{w_{uv}}{\sum_{k \in \text{Out}(u)} w_{uk}} S^{(t)}(u)
\end{equation}
where $d$ is the damping factor (commonly set to 0.85), $In(v)$ denotes the set of all nodes pointing to $v$, $Out(u)$ denotes the set of all nodes pointed to by node $u$, and $w_{uv}$, $w_{uk}$ denote the weights of the respective edges.

We sort the nodes in descending order according to their PageRank scores $S(v)$, and select the top k nodes to obtain the subgraph $Subkg_{PR}$.

\subsection{Subgraph Fusion}
In the subgraph fusion module, we primarily accomplish two tasks: scoring the subgraphs using a query-aware attention reward model, and integrating the other subgraphs into the highest-scoring subgraph.

First, based on the three KG subgraphs constructed using the multi-path algorithm, we employ LLMs to score the dataset in terms of KG richness, question relevance, and connectivity. The scores assigned by the LLMs are used as training data to train query-aware attention reward model (RM). 

To incorporate the semantic information of queries into the scoring process of the RM for KGs, we design a query-aware attention mechanism. We model the attention mechanism between the query vector and the representation of the KG subgraph, thereby emphasizing information regions that are highly relevant to the query semantics and ultimately enhancing the discriminative capability of the reward model.

Specifically, we first obtain the semantic representations of the query input $q \in \mathbb{R}^d $ and the KG subgraph $KGS \in \mathbb{R}^d $. Then, the query and subgraphs representations are linearly projected into the attention space to generate the Query, Key, and Value vectors used in the attention mechanism:
\begin{equation}
\text{Q} = q \cdot W_Q, \text{K} =KGS_i \cdot W_K, \text{V} =KGS_i \cdot W_V
\end{equation}
where, $W_Q \in \mathbb{R}^{d \times d }$, $W_K \in \mathbb{R}^{d \times d }$, and $W_V \in \mathbb{R}^{d \times d }$ denote trainable weight matrices.

We employ a query-aware attention mechanism to compute the query-aware response representation of the subgraph:
\begin{equation}
\text{head} = \text{Softmax}\left( \frac{\text{Q} \text{K}^\top}{\sqrt{d}} \right) \text{V}
\end{equation}
\begin{equation}
\text{Attention}(\text{Q}, \text{K}, \text{V}) = \text{Concat}(\text{head}_1, \ldots, \text{head}_h) W^O
\end{equation}
where $W^O$ denotes trainable weight matrices, $\text{Concat}()$ denotes concatenation function.

This module design ensures that the reward computation not only depends on the inherent representation of the KG subgraph but also reflects the degree of semantic alignment with the query, thereby providing improved contextual adaptability and semantic guidance.

Subsequently, we apply RM to score the three KG subgraphs and select the highest-scoring subgraph. The selection process is described by Eq. (9):
\begin{multline}
KGS_{max} = \max \big( RM\left( KGS_{onehop} \right),\\
RM\left( KGS_{mulhop} \right),\ RM\left( KGS_{PR} \right) \big)
\end{multline}
where max() denotes selecting the highest score from the RM's outputs and returning the subgraph with the top score.

For subgraphs generated from different sources, although some subgraphs may have lower scores, they can still contain important triples highly relevant to the query entity. To fully exploit this potential valuable information, we design a similarity-based subgraph fusion method. We compute the similarity between the highest-scoring subgraph and the query, setting this similarity as a threshold $r$. The specific calculation formula is as follows:

\begin{equation}
r = \cos \left( KGS_{max}, q \right)
\end{equation}

For each lower-scoring subgraph $LS_i$={(h,r,t)}, we calculate the similarity between each triple and the query as follows:

\begin{equation}
\text{sim}\left( \left( h, r, t \right), q \right) = \cos \left( \left( h, r, t \right), q \right)
\end{equation}

If a triple satisfies Eq. (12), it is considered highly relevant to the query and should be retained. All such triples are set as $T_{selected}$.

\begin{equation}
\mathrm{sim}((h,r,t),q) \geq \tau
\end{equation}

Finally, the fused subgraph $G_{fusion}$ is obtained by Eq. (13):
\begin{equation}
G_{fusion} = Subkg_{max} \cup T_{selected}
\end{equation}
\subsection{Subgraph Utilization}
In the process of query-aware retrieval, model performance is often limited by the narrow scope of information coverage and insufficient semantic relevance. To address this issue, we propose a query expansion mechanism enhanced with a KG. It works together with $G_{fusion}$ to improve retrieval relevance and support the performance of RAG. Specifically, we extract an entity set $e_{fusion}$, a relation set $r_{fusion}$, and a triple set $T_{fusion}$ from $G_{fusion}$. The query is concatenated with each entity, relation and triple to form distinct retrieval items, which are then used to perform similarity searches in the vector database. The resulting document sets are aggregated into the final collection $Doc$.

To further enhance the alignment between retrieved passages and the query task, we employ the bge-rerank re-ranking model to score all candidate documents in $Doc$ based on their relevance. The top-$k$ documents are then selected according to the ranking results. This process is formally defined in Eq. (14).

\begin{equation}
C_{rank} = \text{top}\left( rerank(Doc) \right)
\end{equation}

Subsequently, the filtered set $C_{rank}$ is used as contextual input to the LLMs, which summarize and synthesize the information from the selected passages to generate the final answer.

\begin{equation}
\text{Output} = \text{LLM}(C_{rank})
\end{equation}

\subsection{QMKGF algorithmic}
Through the above process, we obtain the response generated by LLMs based on chunks filtered and expanded using the QMKGF algorithm, which incorporates semantic information from KG subgraphs into the original query. The overall algorithm workflow is as shown in Table 1.

\begin{table}[h]
    \centering
    \footnotesize
    \label{tab:_algorithm}
    \caption{The overall QMKGF algorithm.}
    \begin{tabular}{p{8cm}}
        \toprule
        \textbf{Algorithm} QMKGF \\
        \textbf{Input}  Query $q$, Documents chunks, LLMs
        \\
        \textbf{Output}  LLMs output \\
        \midrule
        1.  Begin \\
        2.  Construct KG using LLMs\\
        3.  Fine-tune embedding on domain datasets \\
        4.  $\mathit{document}\_{VB}$  $\leftarrow$ document, $\mathit{ent}\_{VB}$ $\leftarrow$ entity \\
        5.  For $q$ in query sets:\\
        6.  \quad Extract potential entities $e_p$ from the query \\
        7.  \quad $e_t$  $\leftarrow$ mapping of $e_p$ in KG\\
        8.  Construct multiple subgraphs based on one-hop relations $Subkg_{onehop}$, multi-hop relations $Subkg_{mulhop}$, and importance-based relations $Subkg_{PR}$ of $e_t$ \\
        9.  \quad Design a reward model function based on query-aware attention mechanism \\
        10.  \enspace\enspace $\text{Q}$ $\leftarrow$ $q \cdot W_Q$, $\text{K}$ $\leftarrow$ $KGS_i \cdot W_K$, $\text{V}$ $\leftarrow$ $KGS_i \cdot W_V$\\
        11. \enspace\enspace  $\text{head}$ $\leftarrow$ $\text{Softmax}\left( \frac{\text{Q} \text{K}^\top}{\sqrt{d}} \right) \text{V}$ \\
        12. \enspace\enspace  $\text{Attention}(\text{Q}, \text{K}, \text{V})$ $\leftarrow$ $\text{Concat}(\text{head}_1, \ldots, \text{head}_h) W^O$ \\
        13. \enspace\enspace $score$ $\leftarrow$ $linear(\text{Attention})$ \\
        14. \enspace\enspace $KGS_{max}$ $\leftarrow$ $max(RM(KGS_{onehop},KGS_{mulhop},KGS_{PR}))$ \\
        15.  \enspace\enspace  $r$ $\leftarrow$ $\cos \left( KGS_{max}, q \right)$\\
        16. \enspace\enspace $T_{selected} \leftarrow \{(h,r,t)\}$ where $ \mathrm{sim}((h,r,t),q) \geq \tau$ \\
        17.  \enspace\enspace $G_{fusion}$ $\leftarrow$ $Subkg_{max}$  $\cup T_{selected}$\\
        18.  \enspace\enspace Semantically expand the query $q$ based on $G_{fusion}$ \\
        19. \enspace\enspace Retrieve relevant documents $Doc$\\
        20. \enspace\enspace $C_{rank}$ $\leftarrow$ $\text{top}\left( rerank(Doc) \right)$\\
        21. \enspace\enspace Input the top $k$ chunks $C_{rank}$ into the LLMs to generate output\\
        22.  End begin \\
        \bottomrule
    \end{tabular}
\end{table}
\section{Experiment Setup}
In this section, we introduce the datasets, baseline models, experimental settings, and metrics.
\subsection{Datasets}
We used Cultour, IIRC, HotpotQA, SQuAD and MuSiQue datasets to evaluate the models.

\textbf{Cultour} \cite{cultour} is a tourism-oriented dataset comprising 12,000 QA pairs, sourced from both manually curated travel-related materials and content automatically generated by large language models. It is employed to assess model performance in domain-specific tourism scenarios.

\textbf{IIRC} \cite{iirc} is built from English Wikipedia and includes over 13,000 questions. Each question offers only partial context, requiring models to retrieve and synthesize information scattered across multiple related passages. This dataset serves to test comprehension under conditions of incomplete or distributed information.

\textbf{HotpotQA} \cite{hotpotqa} consists of around 113K questions, where correct answers demand the integration of facts drawn from two distinct Wikipedia articles. It is used to benchmark a model’s capability in cross-document reasoning and complex inference.

\textbf{SQuAD} \cite{squad} is a benchmark reading comprehension dataset where answers are typically embedded within a designated paragraph of a Wikipedia entry. It evaluates how well a model can extract and comprehend localized textual information.

\textbf{MuSiQue} \cite{musique} focuses on multi-hop reasoning, featuring questions that generally require 2–4 logical steps to reach a conclusion. It is utilized to assess the model’s capacity for handling multi-hop inference tasks across multiple supporting contexts.

\subsection{Baseline}
In this section, we introduce the baseline models in detail.

\textbf{LLMs-only}: The query is directly input into the LLMs without retrieving any relevant documents from an external database. The final answer is generated solely by the LLMs.

\textbf{W-RAG}: Based on the input query, relevant information is retrieved from the database and appended to the query as supplementary context, which is subsequently processed by the LLMs to produce the final answer.

\textbf{BM25} \cite{bm25} is a statistics-based information retrieval method that scores the relevance between documents and queries by combining the term frequency of query words within a document and the inverse document frequency of those terms across the entire corpus.

\textbf{BGE-rerank} \cite{C-pack} first performs an initial retrieval based on semantic similarity. Subsequently, the BGE-rerank model is applied to re-rank the retrieved results, thereby improving the relevance of the documents input into LLMs.

\textbf{BCE-rerank} \cite{BCEmbedding}, developed by NetEase Youdao, is a bilingual and cross-lingual semantic embedding model specialized in enhancing semantic search accuracy and refining the ranking order of search results according to semantic relevance.

\textbf{LLMs-KG} performs semantic retrieval on an automatically generated KG based on entities extracted from the query, and incorporates the retrieved information into the input of the LLMs to enhance the quality of the generated output.

\textbf{LLMs-KG-rerank} extracts entities from the query and retrieves related chunks from an automatically constructed KG. The retrieved chunks are then re-ranked using the BGE-Rerank model, and the top-ranked chunks is input into the LLMs for answer generation.
\begin{table*}[ht]
    \centering
    \small
    \label{tab:Ablation_experiments}
    \caption{The overall experimental results of QMKGF and other baselines on SQuAD, IIRC, Cultour datasets. The best results are in bold.}
    \begin{tabular}{c|cccc|cccc|cccc}
        \toprule
        \multirow{2}{*}{Models}
        & \multicolumn{4}{c|}{SQuAD} & \multicolumn{4}{c|}{IIRC}& \multicolumn{4}{c}{Cultour}  \\
        \cmidrule(lr){2-5} \cmidrule(lr){6-9}\cmidrule(lr){10-13} 
         & R-1 & R-L & B-1 & Met. & R-1 & R-L & B-1 & Met.& R-1 & R-L & B-1 & Met. \\
        \midrule
         LLMs-only  & 8.17 & 7.96& 20.12 & 14.12& 5.09 & 5.06 & 14.16 & 11.49 & 26.51 & 18.72 &35.95 & 20.09 \\
         W-RAG  & 40.58 & 40.46 & 46.63 & 38.70&45.74 & 45.68 & 52.61 & 37.29 &60.13 &55.68 & 49.78 & 44.57 \\
         BM25  & 43.63 & 43.47 & 48.81 & 41.59&47.27&47.19 &53.04 &38.38 &61.17 &56.44 &50.74 &45.34 \\
         BGE-rerank  &43.47& 43.36 & 48.51 & 41.65& 46.41 & 46.38 & 53.43 &37.42& 61.23 & 56.56 & 50.66 & 45.65\\
         BCE-rerank & 43.25 & 43.13 & 48.73 & 41.67& 45.37 & 45.43 & 52.29&37.16&61.52 & 56.93 & 50.97 & 45.73\\
         LLMs-KG &8.76	&8.53	&19.05	&13.89& 8.09	&8.02&	15.19&	12.20 &30.36&21.67&43.48&23.14\\
         LLMs-KG-rerank  & 43.86	&43.74	&49.44	&41.10& 46.41	&46.26	&54.42	&37.92  &60.44	&55.82	&50.35	&44.64\\
         QCG-rerank  & 45.32	&45.12&	50.95&	42.56& 47.11&	47.05	&55.13&	38.62 &61.89	&57.12	&51.41	&45.49 \\
         QMKGF(ours) & \textbf{51.44}	&\textbf{51.26}&	\textbf{56.26}&	\textbf{47.75} &\textbf{50.37}	&\textbf{50.33}	&\textbf{57.53}&	\textbf{40.38} &\textbf{62.61} & \textbf{58.55} & \textbf{52.12} & \textbf{45.96} \\
        \bottomrule
    \end{tabular}
    
\end{table*}

\textbf{KG2RAG} \cite{kg2rag}  employs a KG-guided chunk expansion process and a KG-based chunk organization process to deliver relevant and important knowledge in well-organized paragraphs.

\textbf{QCG-rerank} \cite{QCG-Rerank} introduces query expansion and chunk graph re-ranking mechanisms. It enriches the semantic representation of the query by extracting and duplicating key information, then constructs a semantic similarity-based chunk graph for iterative re-ranking. The final ranked results are subsequently input into LLMs.

\subsection{Experiment settings}
To ensure experimental fairness, we selected qwen2.5-7B-Instruct as the base LLM and set the temperature to 0.0. During embedding fine-tuning, the number of training epochs was uniformly set to 10, with a maximum query length of 64 tokens and a maximum passage length of 256 tokens. The learning rate was fixed at 1e-5. For the re-ranking stage, we employed the bge-rerank model, utilizing the BGE Large embedding for vector representations. For the reward model (RM), we used ernie-3.0-base with a learning rate of 1e-5, a maximum input length of 512 tokens, and trained for 10 epochs. Entity and relation extraction as well as KG construction for Chinese datasets were performed using the qwen2.5 model, while for English datasets, Llama 3 was used. For the IIRC dataset, 4,906 samples were used for training and 593 for testing. The Cultour dataset was split into training and testing sets with an 8:2 ratio. For HotpotQA, SQuAD, and MuSiQue datasets, the same 8:2 split was applied. If the test set contains more than 1,000 samples, a random subset of 1,000 samples is selected for evaluation. All experiments were conducted on an NVIDIA RTX A6000 GPU. 

\subsection{Metrics}
To evaluate model performance, we adopt three metrics: ROUGE, BLEU, and METEOR. ROUGE \cite{rouge} captures lexical overlap between generated outputs and human-annotated references, emphasizing recall. BLEU \cite{bleu} focuses on precision by computing n-gram co-occurrence between candidate and reference sequences. METEOR \cite{meteor} provides a more fine-grained evaluation by accounting for not only exact word matches but also stemming, synonymy, and word order, thus offering a more comprehensive assessment of generation quality compared to BLEU.

\section{Results}
In this section, we evaluate the proposed QMKGF model through a comparative analysis against a number of representative baselines. Subsequently, ablation studies are conducted to investigate the impact of key components and design choices within our framework.
\begin{table*}[ht]
    \centering
    \small
    \label{tab:Ablation_experiments}
    \caption{The overall experimental results of QMKGF and other baselines on HotpotQA and MuSiQue datasets. The best results are in bold.}
    \begin{tabular}{c|cccc|cccc}
        \toprule
        \multirow{2}{*}{Models}
        & \multicolumn{4}{c|}{HotpotQA} & \multicolumn{4}{c}{MuSiQue}  \\
        \cmidrule(lr){2-5} \cmidrule(lr){6-9} 
         & R-1 & R-L & B-1 & Met. & R-1 & R-L & B-1 & Met. \\
        \midrule
         LLMs-only & 11.86	&11.73	&25.38	&17.91 & 2.49	&2.43	&13.20&	10.75  \\
         W-RAG &54.07&	54.01&	60.22	&47.20 &36.23	&36.16	&42.82&	34.89  \\
         BM25 &54.86 &	54.73 &	60.76	 &50.06 &38.53 &	38.20	 &43.64 &	36.36  \\
         BGE-rerank & 55.26	 &55.15	 &61.30	 &49.91 &39.51	 &39.43	 &45.50	 &37.05 \\
         BCE-rerank&54.97	 &54.87 &	60.45 &	49.43 & 39.13	 &39.11	 &45.37	 &36.83 \\
         LLMs-KG&12.59	&12.52	&25.11	&17.77  & 2.89&	2.82	&11.70	&9.99 \\
         LLMs-KG-rerank &54.16	&54.06	&60.04	&48.50 & 38.82	&38.73	&44.64	&36.08 \\
         QCG-rerank&57.44 & 	57.24 & 	62.69 & 	51.61 & 41.17 & 	40.98 & 	46.53 & 	38.31 \\
         KG2RAG&58.90&	58.83&	66.65&	53.32 & 37.41&	37.09	&48.25	&35.71 \\
         QMKGF(ours)&\textbf{64.98}	&\textbf{64.95}	&\textbf{68.42}	&\textbf{57.74} & \textbf{47.42}	&\textbf{47.35}	&\textbf{53.31}&\textbf{43.95}  \\
        \bottomrule
    \end{tabular}
    
\end{table*}
\subsection{Main results}
We evaluated QMKGF on the SQuAD, IIRC, Cultour, HotpotQA, and MuSiQue datasets, and the detailed results are presented in Table 2 and Table 3.

As shown in Tables 2 and 3, our proposed QMKGF framework consistently outperforms all baseline models across five benchmark datasets: SQuAD, IIRC, Cultour, HotpotQA, and MuSiQue, with the best performance highlighted in bold. In particular, QMKGF achieves the highest scores in nearly all evaluation metrics, such as ROUGE-1, ROUGE-L, BLEU-1, and METEOR. The most notable gain is on the HotpotQA dataset, where QMKGF surpasses bge-rerank by 9.72. QMKGF attains a remarkable 64.98 ROUGE-1 and 64.95 ROUGE-L, outperforming the previous best (KG2RAG) by over 6.0 points. Similar results were achieved on two other multi-hop question answering datasets, IIRC and MuSiQue, demonstrating its superiority in multi-hop reasoning tasks. QMKGF achieves significant improvements on the Chinese question answering dataset Cultour. Specifically, it outperforms the rerank-only baseline bge-rerank by 1.39 points on the ROUGE-1 metric. On the English extraction dataset SQuAD, QMKGF is 7.97 points higher than the BGE-rerank model in the ROUGE-1 metric. This demonstrates the model’s superiority in single-hop question answering tasks. 

Moreover, across all datasets, QMKGF consistently outperforms both the LLMs+KG and LLMs+KG+rerank. These results indicate that QMKGF effectively enhances the quantity of relevant chunks by leveraging query-aware attention RM to filter and integrate multiple subgraph paths. The reranked results are then passed to LLMs, leading to improvements in both the relevance and reliability of the generated responses.

\subsection{Ablation experiment}
To assess the effectiveness of our proposed approach, experiments were conducted on the Chinese Cultour dataset and the English HotpotQA dataset. The main tests focused on the impact of the following components on the model's performance: without query-aware attention (w/o-attention), and without fine-tuned embedding (w/o-fintune). The detailed results are shown in Table 4.

\begin{table*}[h]
    \centering
    \small
    \label{tab:Ablation_experiments}
    \caption{Ablation experiments of QMKGF. The best results are in bold.}
    \begin{tabular}{c|cccc|cccc}
        \toprule
        \multirow{2}{*}{Models}
        & \multicolumn{4}{c|}{HotpotQA} & \multicolumn{4}{c}{Cultour}  \\
        \cmidrule(lr){2-5} \cmidrule(lr){6-9} 
         & R-1 & R-L & B-1 & Met. & R-1 & R-L & B-1 & Met. \\
        \midrule
         Bge-rerank  & 55.26 & 55.15 & 61.30 & 49.91 & 61.23 & 56.56 & 50.66 & 45.65 \\
         w/o-attention & 61.17 & 61.13 & 65.79 &54.57&61.76 & 58.03 & 49.86 & 45.10\\
         w/o-fintune & 61.20 & 61.15 & 65.83 &54.90&61.77 & 58.06 & 49.69 & 45.13\\
         QMKGF & \textbf{64.98} &\textbf{ 64.95} & \textbf{68.42} &57.74&\textbf{62.61} & \textbf{58.55} & \textbf{52.12} & 45.96  \\
        \bottomrule
    \end{tabular}
    
\end{table*}

As shown in Table 4, each component contributes positively to the final performance. QMKGF achieves the best performance on R-1, R-L, and B-1, demonstrating the effectiveness of the proposed multi-path KG subgraph construction strategy. In the results on the English dataset HotpotQA, we observe that all components of our framework lead to a substantial improvement over the BGE-rerank baseline. While similar gains are also present on the Chinese dataset, the improvement is less pronounced compared to IIRC. Additionally, without personalized PageRank algorithm has a more significant negative impact on English datasets than on the Chinese dataset. One possible reason is that we adopt the Qwen2.5 model as the backbone, which has been extensively pretrained on Chinese corpora, leading to stronger prior knowledge in the Chinese domain. In contrast, its relative lack of exposure to English knowledge makes the inclusion of high-quality triples from diverse subgraphs more influential for final performance on English tasks.

The performance impact of removing query-aware attention and removing the fine-tuned embedding shows a similar trend. These results suggest that QMKGF effectively integrates one-hop relations, multi-hop relations, and the importance-based relations algorithm to construct and fuse multiple semantic paths. In the process of scoring the three subgraphs using the query-aware attention reward model, the influence of the query on each subgraph is taken into account. Thereby improving the quality of the retrieved documents and enhancing the performance of LLMs in response generation.

\subsection{Impact of fine-tuning embeddings}
To evaluate the impact of fine-tuned embedding models on retrieval performance, we conducted experiments using metrics such as MRR@1, MRR@10, Recall@10, and nDCG@10. As illustrated in Fig. 4, the fine-tuned models show consistent and substantial improvements across all metrics on both English datasets (SQuAD, HotpotQA) and the Chinese dataset (Cultour). For example, on the SQuAD dataset, fine-tuning the bge\_base model results in a 9.27-point increase in Recall@10, while the bge\_large model achieves an even greater improvement of 12.82 points. This indicates that fine-tuning the bge\_large model has a greater impact on the final performance. Moreover, we observe that the performance improvement from embedding fine-tuning is more pronounced on the English datasets compared to the Chinese dataset. This is likely because the Chinese dataset already achieved relatively strong performance before fine-tuning, leaving limited room for further improvement, which in turn makes the fine-tuning gains less noticeable. These results demonstrate that domain-specific embedding fine-tuning can effectively enhance the accuracy of relevant document retrieval.

\begin{figure*}[ht]
    \centering
    \begin{subfigure}[b]{0.32\textwidth}
        \centering
        \includegraphics[width=\textwidth]{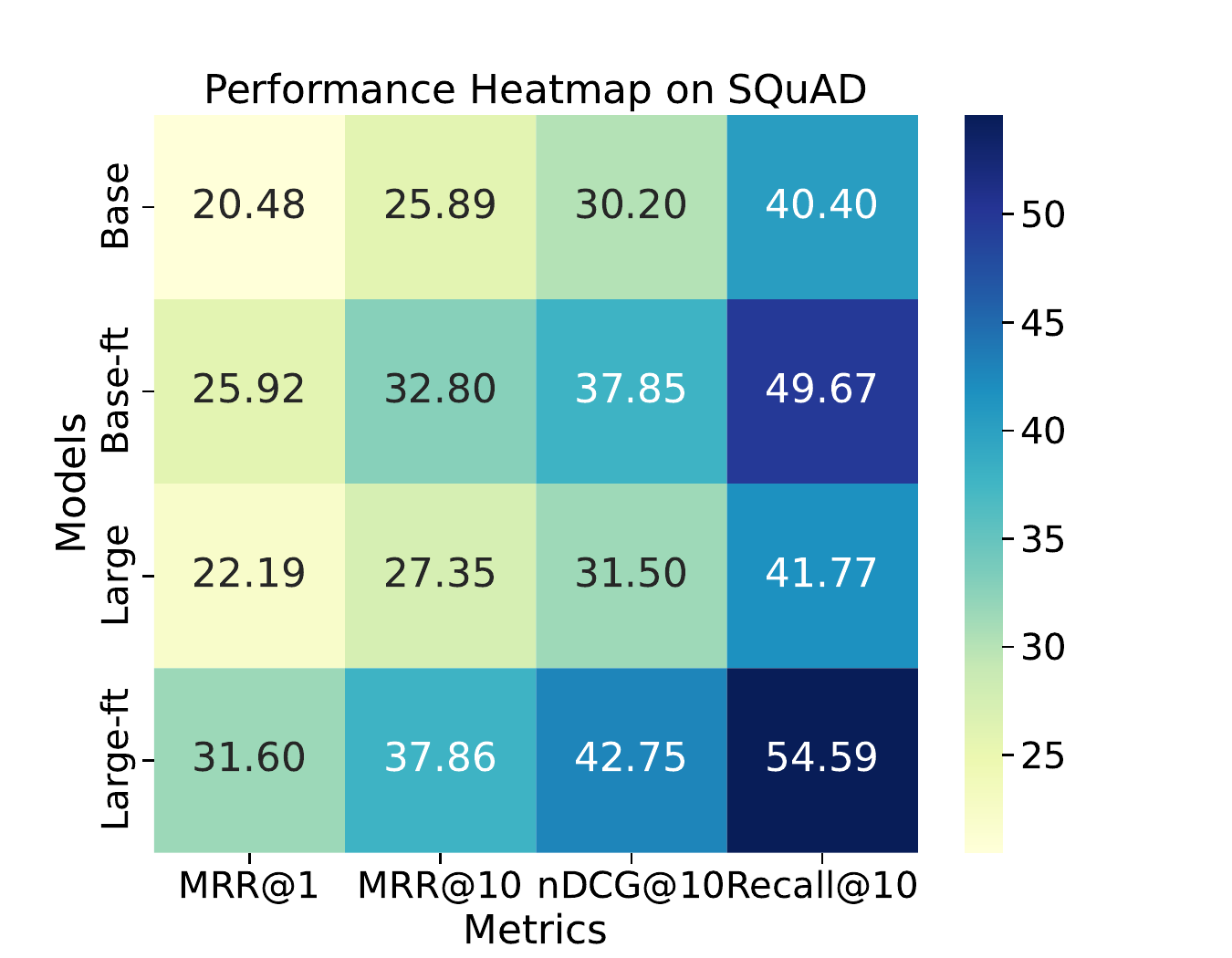}
        \caption{Results on SQuAD}
        \label{fig:5a}
    \end{subfigure}
    \begin{subfigure}[b]{0.32\textwidth}
        \centering
        \includegraphics[width=\textwidth]{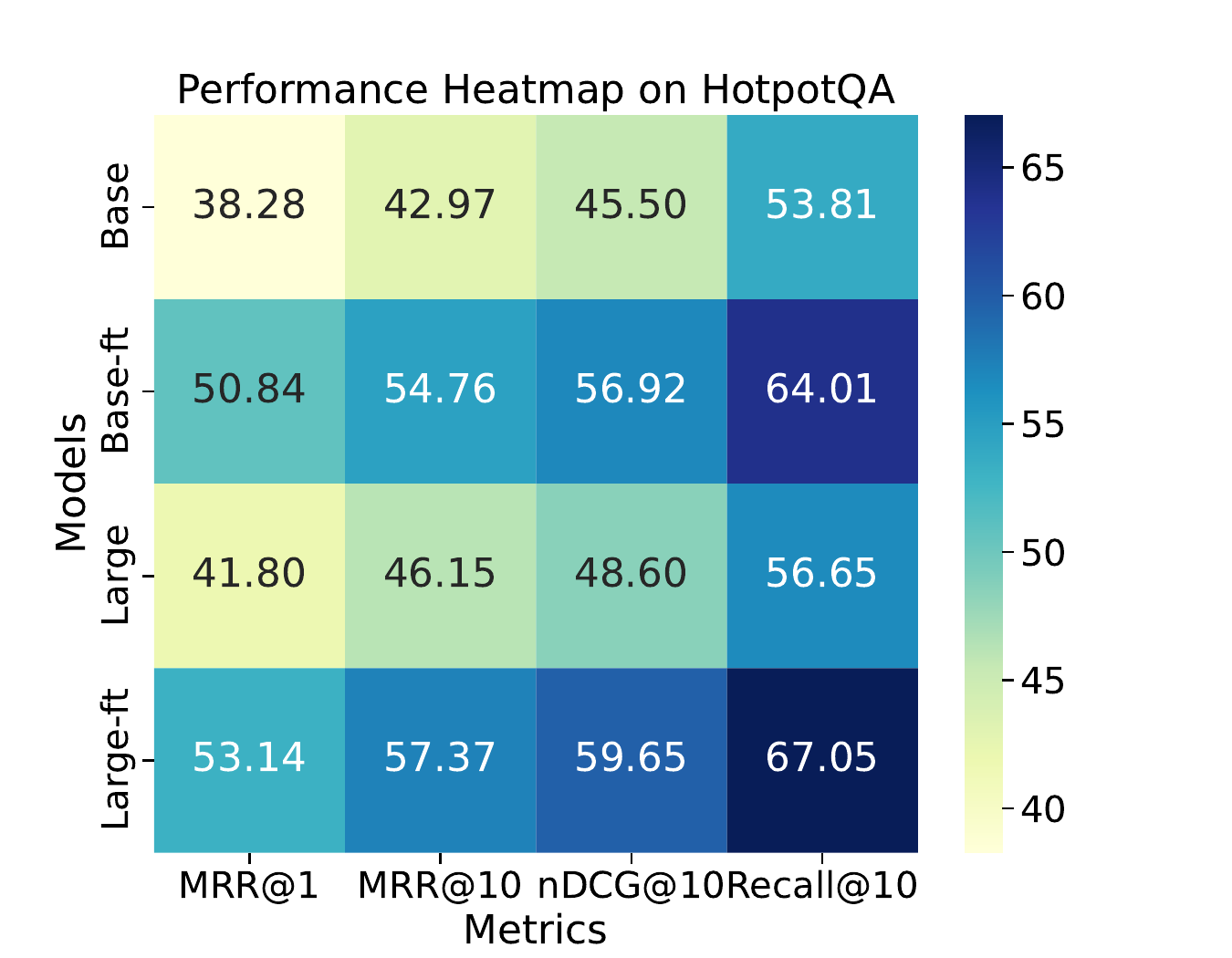}
        \caption{Results on HotpotQA}
        \label{fig:5b}
    \end{subfigure}
    \begin{subfigure}[b]{0.32\textwidth}
        \centering
        \includegraphics[width=\textwidth]{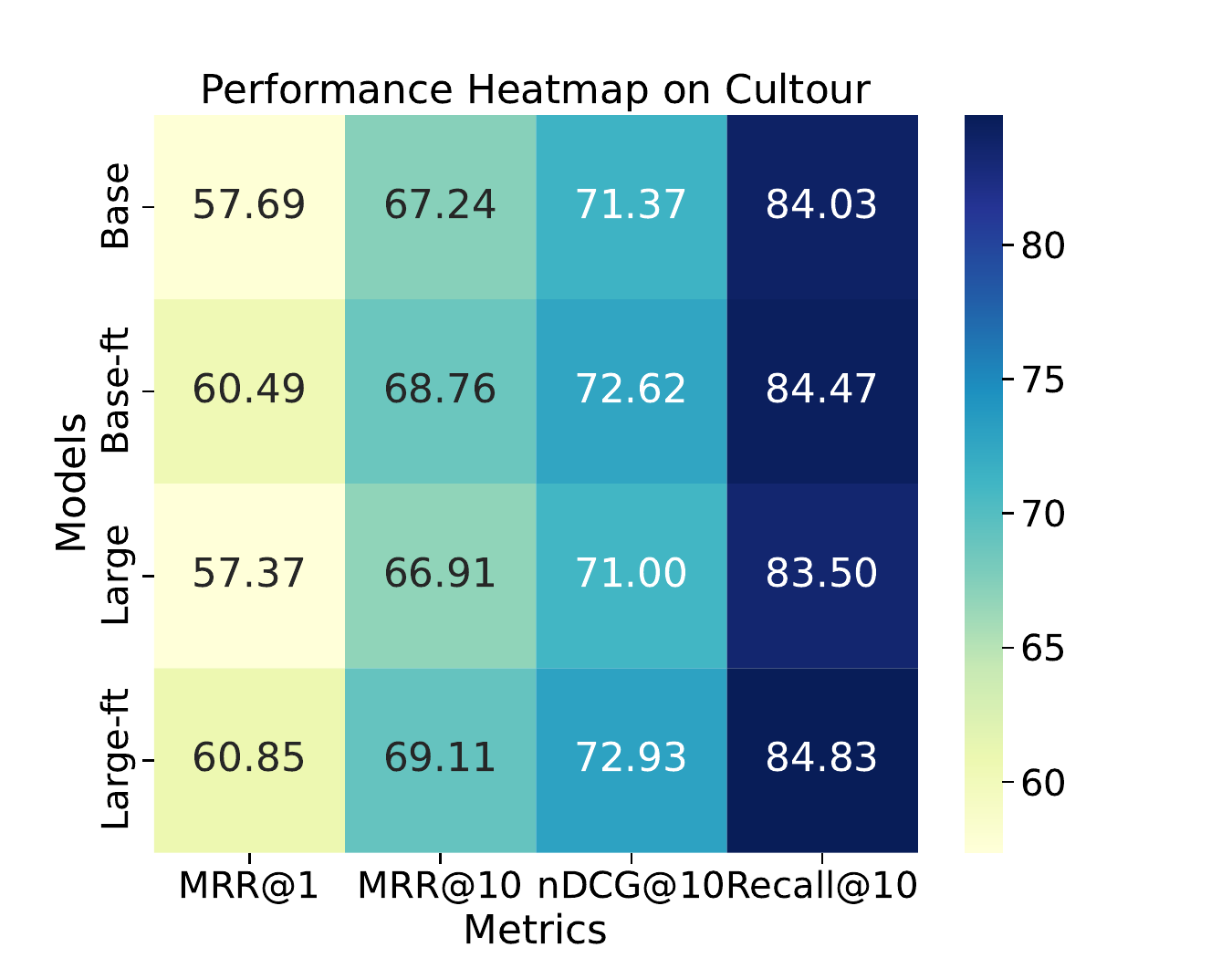}
        \caption{Results on Cultour}
        \label{fig:5b}
    \end{subfigure}
    \caption{Impact of fine-tuning embeddings (combined view).}
    \label{fig:5}
\end{figure*}

\subsection{Assessment of Information Retrieval Ability}
In addition to evaluating text quality using ROUGE, BLEU, and METEOR, we also incorporate Hit@10, as well as Precision, Recall, and F1-score, to provide a complementary assessment of QMKGF’s generated results from the perspectives of answer matching and information retrieval. These metrics offer a more comprehensive evaluation of the relevance and accuracy of the generated content. 

\begin{table}[h]
    \centering
    \small
    \label{tab:fine-tuning embeddings}
    \caption{Results on Precision, Recall, and F1-score. The best results are in bold.}
    \begin{tabular}{c|cccc}
        \toprule
        \multirow{2}{*}{Models}
        & \multicolumn{4}{c}{Cultour}   \\
        \cmidrule(lr){2-5}  
          & P & R & F1 & hit@10 \\
        \midrule
         W-RAG 	&63.44	&\textbf{53.61}	&58.12&97.70\\
         BGE-rerank 	&62.03&	53.43	&57.41&98.00\\
         QMKGF(ours)	&\textbf{71.60}&	51.53	&\textbf{59.24} & \textbf{99.40}\\
        \bottomrule
    \end{tabular}
    
\end{table}

As shown in Table 5, QMKGF outperforms both the W-RAG and BGE-rerank models. In the hit@10 metric, QMKGF achieved 99.4\%, reaching the highest performance, indicating that our model has a significant advantage in the accuracy of answer retrieval and matching. In terms of Precision (P) and F1-score(F1), our model achieved the best performance, demonstrating excellent accuracy in generating results. The results indicate that our multi-path knowledge graph approach effectively expands the semantic neighborhood of query entities, broadens the recall scope, and enhances query relevance, leading to notable improvements in retrieval and generation performance.

\subsection{Impact of reward model}
To further evaluate the impact of using different pretrained models as reward models in QMKGF, we conducted comparative experiments with BERT \cite{BERT} and ERNIE \cite{ernie}. Both models possess strong semantic understanding capabilities and have been widely adopted in various NLP tasks. As shown in Table 6, employing ERNIE as the RM for subgraph selection leads to greater improvements in the accuracy of the LLM’s outputs, demonstrating its superior effectiveness in guiding the retrieval process.
\begin{table*}[h]
    \centering
    \small
    \label{tab:fine-tuning embeddings}
    \caption{The impact of reward model. The best results are in bold.}
    \begin{tabular}{c|cccc|cccc}
        \toprule
        \multirow{2}{*}{Models}
        & \multicolumn{4}{c|}{Cultour} & \multicolumn{4}{c}{SQuAD}  \\
        \cmidrule(lr){2-5} \cmidrule(lr){6-9} 
         & R-1 & R-L & B-1 & Met. & R-1 & R-L & B-1 & Met. \\
        \midrule
         ERNIE & \textbf{62.61} &	\textbf{58.55}	 &\textbf{52.12} &\textbf{45.96} & \textbf{51.44}	 &\textbf{51.26} &	\textbf{56.26}	 &\textbf{47.75}  \\
         BERT &62.06	 &58.40	 &49.97	 &45.30 &51.05 &	50.86 &	55.89	 &47.45 \\
        \bottomrule
    \end{tabular}
    
\end{table*}

These improvements can be attributed to the semantic knowledge integration mechanisms of ERNIE, which leverages entity-level masking and prior knowledge during pretraining. Unlike BERT, which is limited to token-level representations, ERNIE incorporates structured knowledge from knowledge graphs and factual information during its representation learning process. This enables the reward model to better evaluate the semantic relevance between the candidate triples and the query, thereby guiding the selection of more informative subgraphs and enhancing the overall output quality.
\begin{figure*}[h]
	\begin{center}
		\includegraphics[scale=0.7]{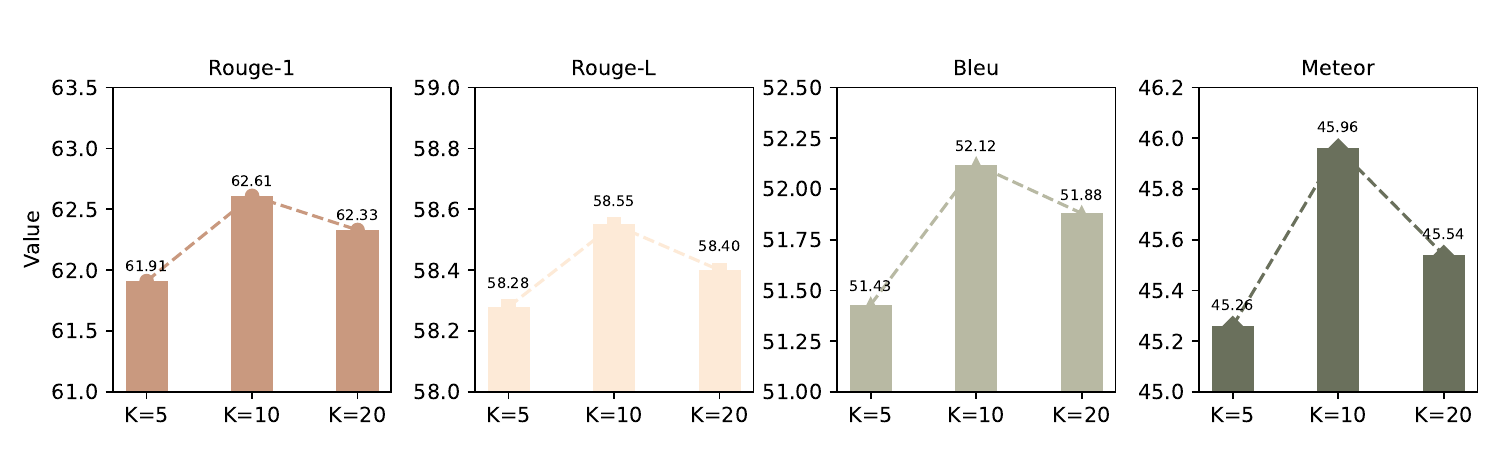}
		\caption{Effect of subgraph node count.}
		\label{fig:1}       
	\end{center}
\end{figure*}

\subsection{Effect of Subgraph Node Count}

To evaluate the impact of different subgraph sizes on the generation performance of QMKGF, we conducted comparative experiments on Cultour datasets with three configurations, setting the number of subgraph nodes $K$ to 5, 10, and 20, respectively.

As shown in Fig. 5, the model achieves optimal performance when $K$=10, with the highest scores across all evaluation metrics, including ROUGE-1 (62.61) and BLEU (52.12). This indicates that the configuration with $K$=10 strikes a favorable balance between semantic relevance and information redundancy, thereby effectively enhancing generation quality. When the number of nodes is smaller ($K$=5), the performance on ROUGE metrics remains relatively close, but BLEU and METEOR scores are notably lower. This suggests insufficient semantic coverage in the subgraph, which results in less informative and under-supported generated content. On the other hand, increasing the node count to $K$=20 leads to a slight decline in performance, particularly in BLEU and METEOR scores. This decline may be attributed to the introduction of irrelevant or noisy entities and relations, which interferes with semantic alignment and coherent context construction during generation.

\subsection{The impact of head nums on RM performance}
Fig. 6 illustrates the impact of the number of attention heads in the multi-head attention mechanism on model performance, showing notable differences across the HotpotQA, Cultour, and SQuAD datasets. Experimental results demonstrate that increasing the number of attention heads significantly improves performance on HotpotQA, with the highest scores observed at 32 and 64 heads. This suggests that for complex multi-hop reasoning tasks, introducing more parallel attention paths facilitates the capture of diverse information sources and the modeling of long-range dependencies, thereby enhancing the model's ability to identify semantically relevant knowledge paths. In contrast, the performance improvements on Cultour and SQuAD are relatively modest. This may be attributed to the relatively lower reasoning complexity and simpler information structures of these tasks, where an excessive number of attention heads could introduce redundant computations or semantic noise, thus weakening the model's focus on core information. 

\begin{figure}[h]
	\begin{center}
		\includegraphics[scale=0.35]{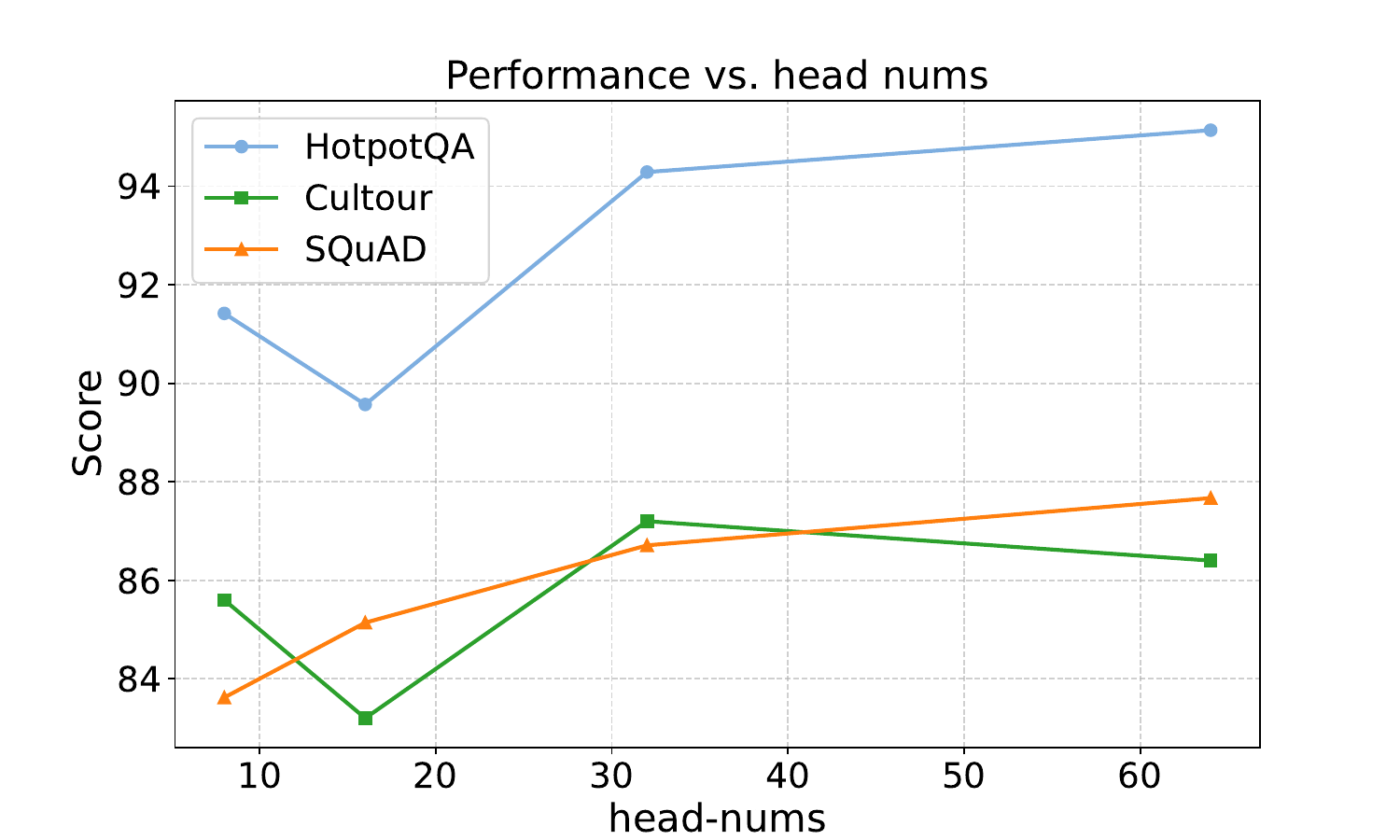}
		\caption{The impact of head nums on RM performance.}
		\label{fig:1}       
	\end{center}
\end{figure}

These results indicate that increasing the number of attention heads is highly beneficial for complex tasks such as multi-hop question answering, as it enhances the model's capacity to capture and integrate knowledge paths. However, for tasks with more concentrated and structurally straightforward information, a moderate number of attention heads may already suffice for high-quality modeling, and further increases could lead to performance instability.

\subsection{Impact of different subgraph fusion strategies}
To further examine the impact of different subgraph fusion strategies on model performance, we conduct experiments comparing three representative approaches—All fusion, Top-5 triples fusion, and RM fusion—on the HotpotQA and MuSiQue datasets. All Fusion: All triples from each subgraph are retained without any filtering. Top-5 Triples Fusion: Only the top 5 triples from each subgraph are preserved to reduce noise. RM Fusion: A reward model is used to score and filter triples, retaining only those highly relevant to the query to enhance information quality and generation performance. The generation quality is evaluated using four standard metrics: ROUGE-1 (R-1), ROUGE-L (R-L), BLEU-1 (B-1), and METEOR (Met.). The experimental results are shown in Table 7. 

\begin{table*}[h]
    \centering
    \small
    \label{tab:fine-tuning embeddings}
    \caption{Impact of different subgraph fusion strategies. The best results are in bold.}
    \begin{tabular}{c|cccc|cccc}
        \toprule
        \multirow{2}{*}{Models}
        & \multicolumn{4}{c|}{HotpotQA} & \multicolumn{4}{c}{MuSiQue}  \\
        \cmidrule(lr){2-5} \cmidrule(lr){6-9} 
         & R-1 & R-L & B-1 & Met. & R-1 & R-L & B-1 & Met. \\
        \midrule
         All fusion &64.07	&64.05	&67.62	&57.01	&46.36	&46.28	&52.42	&43.24 \\
         Top-5 triples fusion &63.54	&63.51	&67.26	&56.84	&43.33	&44.24	&50.93	&41.67 \\
         RM fusion &\textbf{64.98}&	\textbf{64.95}&\textbf{68.42}&\textbf{57.74}&	\textbf{47.42}&	\textbf{47.35}	&\textbf{53.31}	&\textbf{43.95}\\
        \bottomrule
    \end{tabular}
    
\end{table*}

\begin{figure*}[h]
	\begin{center}
		\includegraphics[scale=0.6]{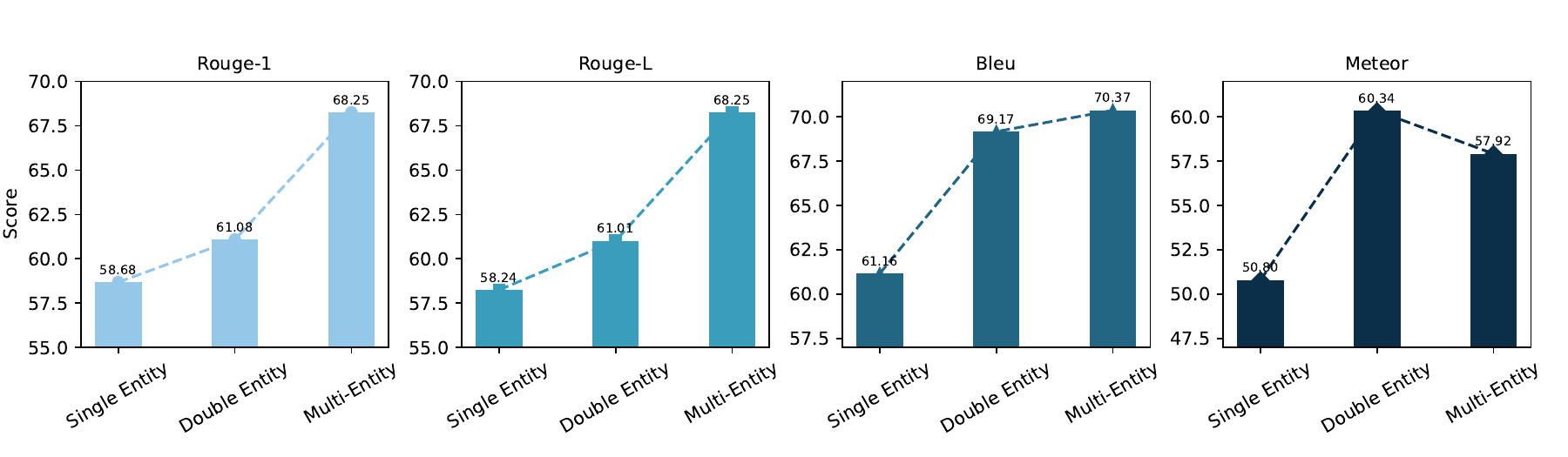}
		\caption{Effect of subgraph node count.}
		\label{fig:1}       
	\end{center}
\end{figure*}
Different subgraph fusion methods have a significant impact on the quality of the final generated results. Notably, the Top 5 triples fusion method performs well on the HotpotQA dataset but poorly on the MuSiQue dataset. The RM fusion method achieves the best results on both HotpotQA and MuSiQue, demonstrating that this strategy has a certain degree of generalizability. This is because All fusion may introduce excessive redundant or noisy information, which interferes with the model’s focus on key content. Top-5 triples fusion reduces noise to some extent, but its fusion strategy is relatively coarse. It fails to effectively model semantic correlations, resulting in fewer relevant triples and thus limited document retrieval. In contrast, RM fusion models semantic relevance across multiple heterogeneous subgraphs more effectively. This helps retain the most valuable information, enables the retrieval of more relevant documents, and ultimately improves the quality of generated content.

\subsection{Impact of entity quantity in queries}

To further investigate the impact of entity quantity in queries on model performance, we divide the HotpotQA test set into three categories: single-entity queries, two-entity queries, and multi-entity queries (more than 2 entities). For each category, 200 samples are randomly selected for evaluation. The experimental results are shown in Fig. 7. The experimental results demonstrate a clear upward trend in the scores of the QMKGF model across three evaluation metrics—ROUGE-L, ROUGE-1, and BLEU—as the number of entities in the query increases, indicating enhanced generation capabilities. For instance, under the BLEU metric, the score improves significantly from 61.16\% in the single-entity group to 70.37\% in the multi-entity group.

This phenomenon reflects that, in the context of the multi-hop question answering task represented by HotpotQA, multi-entity queries provide more explicit semantic anchors, which facilitate richer semantic expansion when integrated with knowledge graphs. Consequently, the model is able to retrieve more query-relevant passages from the database during the retrieval stage, offering stronger contextual support for generation. In contrast, single-entity queries often exhibit semantic ambiguity, which may lead to overly broad retrieval scopes or insufficient contextual aggregation, thereby undermining the accuracy and consistency of the generated content.

\section{Conclusion}
In this paper, we proposed QMKGF, a novel framework aimed at enhancing the performance of large language models (LLMs). To capture semantically rich structures associated with the query, a multi-path subgraph construction strategy was developed. Subsequently, a query-aware attention reward model was introduced to score triples based on their semantic relevance to the query, guiding the selection of high-quality triples for knowledge graph fusion. Finally, the updated subgraph was utilized to expand the original query, thereby improving the relevance of retrieved documents. The proposed QMKGF was evaluated on multiple benchmark datasets, including SQuAD, IIRC, Culture, HotpotQA, and MuSiQue, and the experimental results demonstrate its effectiveness. Detailed ablation studies further validate the contributions of the domain-adaptation KG construction module, the query-aware attention reward model, and the subgraph fusion module. Additionally, we investigated the impact of fine-tuned embedding models on overall performance. 
Different pretrained models were evaluated as the backbone of the reward model. In addition, the impact of varying the number of heads in the query-aware attention mechanism on the final performance was also explored. Finally, we analyzed the influence of subgraph size in the knowledge graph on model performance.

Future work will explore the construction of more domain-specific knowledge graphs tailored to the unique characteristics of specialized data. Based on these KGs, the integration of causal relationship modeling and event chain

\bibliographystyle{cas-model2-names}

\bibliography{mybibfile}

\end{document}